\documentclass[conference]{IEEEtran}
\usepackage{graphicx}
\usepackage{amsmath}
\usepackage{mathtools}
\usepackage{latexsym}

\usepackage{amsthm}
\theoremstyle{definition}
\newtheorem{theorem}{Definition}
\newtheorem*{theorem*}{Definition}
\newtheorem{definition}[theorem]{Definition}
\newtheorem*{definition*}{Definition}

\usepackage{listings}

\usepackage{algorithm}
\usepackage{algorithmic}

\begin{document}

\title{An experiment of the complexity of sliding block puzzles by 2D heat flow in paramodulation}

\author{

\IEEEauthorblockN{Ruo Ando}
\IEEEauthorblockA{
National Institute of Informatics\\
2-1-2 Hitotsubashi, Chiyoda-ku, Tokyo \\101-8430 Japan\\
}
\and
\IEEEauthorblockN{Yoshiyasu Takefuji}
\IEEEauthorblockA{Musashino University\\
3-3-3 Ariake, Koto-Ku, Tokyo \\
135-8181, Japan
}
}

\maketitle

\begin{abstract}
In this paper, we present a curious experiment with the hot list strategy in solving sliding block puzzles by paramodulation. The hot list strategy is one of the look-ahead strategies using paramodulation in automated reasoning. We define two heat flows in the reasoning process - vertical with the hot list of permutations along the Y-axis and horizontal along the X-axis. In the experiment, we have generated 500 * 8 puzzles under the test of the solvability checking by counting inversions. We have obtained curious 2D and 3D plots of the complexity by defining heat flow with hot lists. We can distinguish a few groups in 500 boards (puzzles) based on the concept of heat-resisting.
\end{abstract}

\IEEEpeerreviewmaketitle

\section{Introduction}

A sliding puzzle (also called a sliding block puzzle) is a combination puzzle where a player slides pieces along certain routes on a board to reach a certain end configuration (state).


\begin{figure}[ht]
\centering
\includegraphics[scale=0.5]{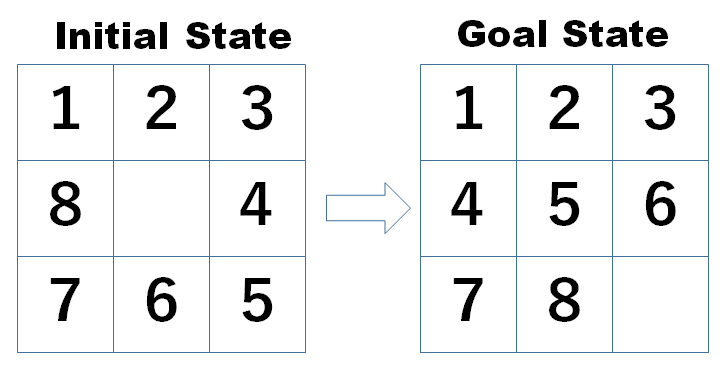}
\caption{Initial state and goal state of 8 puzzle.}
\end{figure}

In sliding puzzles, a player is prohibited from lifting any piece off the board. This constraint separates sliding puzzles from rearrangement puzzles.
Consequently, discovering routes opened up by each move with the two-dimensional confines of the board is an interesting point of solving sliding block puzzles.

Figure 1 shows the example of a sliding puzzle. The puzzle has 9 square slots on a square board.
The first eight slots have square pieces. The 9th slot is empty.

Sliding block can be represented as the permutation.
A permutation of a set S is a bijection from S onto itself. If the set we permuting is $ A = {1,2, ..., n} $, it is often convenient
to represent a permutation $ \sigma $ as follows:

\begin{eqnarray}
\sigma = \left\{
\begin{array}{lllll}
1,&2,&3,& ... &\\
\sigma(1),&\sigma(2),&\sigma(3),& ... ,& \sigma(n)
\end{array}
\right\}
\end{eqnarray}

For instance, consider the set A = {1,2,3,4,5,6}. Then the permutation $\pi$,

\begin{eqnarray}
\pi = \left\{
\begin{array}{llllll}
1,&2,&3,&4,&5,&6\\
4,&1,&5,&2,&3,&6
\end{array}
\right\}
\end{eqnarray}

sends 1 to 4, 2 to 1, 3 to 5 and fixes, or leaves unchanged, the element 6.

The theorem prover OTTER (Organized Techniques for Theorem-proving and Effective Research) has been developed by W. McCune as a product of Argonne National Laboratory.
OTTER is based on earlier work by E. Lusk, R. Overbeek, and others \cite{Lusk2}.
OTTER adopts the given-clause algorithm and implements the set of support strategy \cite{Wos2}.
In this paper we use OTTER for our experiments.

\section{Given Clause Algorithm}

OTTER adopts given-clause algorithm in which the program attempts to use any and all combinations from axioms in the given clause.
In other words, the combinations of the clause are generated from given clauses which have been focused on.
Given clause algorithm is shown in Algorithm 1.

\begin{algorithm}
\caption{Given clause algorithm}
\label{alg1}
\begin{algorithmic}[1]
\renewcommand{\algorithmicrequire}{\textbf{Input:}}
\renewcommand{\algorithmicensure}{\textbf{Output:}}
\REQUIRE SOS, Usable List
\ENSURE Proof
\WHILE{until SoS is empty}
\STATE choose a given clause G from SoS;
\STATE move the clause g to Usable List;
\WHILE { c\_1, ..., c\_n in Usable List}
\WHILE{$ R(c_1,..c_i, G ,c_{i+1},..c_n) exists $}
\STATE $ A \Leftarrow R(c_1,..c_i, G, c_{i+1},..c_n); $
\IF{A is the goal}
\STATE report the proof;
\STATE stop
\ELSE[A is new odd]
\STATE add A to SoS X
\ENDIF
\ENDWHILE
\ENDWHILE
\ENDWHILE
\end{algorithmic}
\end{algorithm}

At line 2, given clause G is extracted from SoS (Set of Support).
Line 4 and 5 is a loop to use any and all combinations of the given clause and Usable List.
In detail, \cite{Slaney} discuss the basic framework of the given clause algorithm.

\section{Paramodulation}

\subsection{Formulation}

Paramodulation, which is introduced by \cite{para}, is a powerful method of equational reasoning.
Paramodulation takes two clauses of which at least cone contains positive equality
literal $l \equiv r$.

\begin{eqnarray}
{
\begin{array}{lllll}
\{L_1, L_2, ... L_n\} \\
\{l \equiv r, K_2 ... K_m\} \exists p, \sigma. \sigma = mgu(L1/p,l) \\ \hline
\{L_1/p \leftarrow r, L_2, ..., L_n, K_2, ..., K_m \} \sigma
\end{array}
}
\end{eqnarray}

According to \cite{graf},
paramodulation is a realization of Leibniz's substitution of equals by equals.
In the case that $L_1$ contains a subterm at a position $p$ which is unifiable with $l$, the paramount
can be computed with the unifier $ \sigma$.

For example, consider two clauses.
\begin{eqnarray}
{
\begin{array}{lllll}
\{P(\gamma, h(f(\alpha,y),\beta))\} \\
\{f(x,\gamma) \equiv g(x), Q(x)\}
\end{array}
}
\end{eqnarray}

(4) generates the new clause.
\begin{eqnarray}
{
\begin{array}{lllll}
\{P(\gamma, h(g(\alpha,\beta)),Q(\alpha)\}
\end{array}
}
\end{eqnarray}

From the mathematical rules $x+0=x$ and $-y+y=0$,
\begin{eqnarray}
{
\begin{array}{lllll}
\{plus(x,0) \equiv x \} \\
\{plus(minus(y),y) \equiv 0 \}
\end{array}
}
\end{eqnarray}
generates from the first into the second of the clauses.
\begin{eqnarray}
{
\begin{array}{lllll}
\{minus(0) \equiv 0 \}
\end{array}
}
\end{eqnarray}

Paramodulation uses unification, while demodulation adopts matching.

\subsection{Implementation}

Concerning the implementation of OTTER, in paramodulation, two parents and a child are processed.
The parent clauses contain the equality applied for the replacement. The parent clauses are divided into two: from parent and from clause.
If equality comes from the literal, the side of equality unifies with the term, which is replaced with from the term.
The replaced term is called the into the term. The literal containing the replaced term is also called the into literal.
Also, the parent containing the replaced term is called the into the parent or into clause.

Paramodulation is divided into two procedures: para\_into and para\_from.

\begin{enumerate}
\item para\_into. Paramodulation into the given clause. When we make an inference by the para\_into rule, we paramodulte into the given clause from containing positive equality and on the usable list.
\item para\_from. Paramodulation from the given clause. When we make an inference by the para\_from rule, the given clause contains positive equality, and the inference is made by paramodulating using this equality into a clause.
\end{enumerate}

In this paper, we use the rule of para\_into. The procedure of para\_into is invoked from infer\_and\_process taking the given clause.

%

\begin{algorithm}
\caption{para\_into (paramodulation into given clause)}
\label{alg1}
\begin{algorithmic}[1]
\renewcommand{\algorithmicrequire}{\textbf{Input:}}
\renewcommand{\algorithmicensure}{\textbf{Output:}}
\REQUIRE given clause
\ENSURE subterm list
\STATE into\_literal = given\_clause $\rightarrow$ first\_literal
\WHILE{into\_literal != NULL}
\STATE subterm\_list = into\_literal $\rightarrow$ term $\rightarrow$ list
\WHILE {subterm\_list != NULL}
\STATE subterm\_list $\rightarrow$ path = 1
\STATE para\_into\_terms(subterm\_list, into\_literal)
\STATE subterm\_list $\rightarrow$ path = 0
\STATE subterm\_list = subterm\_list $\rightarrow$ next
\STATE into\_literal = into\_literal
\ENDWHILE
\ENDWHILE
\end{algorithmic}
\end{algorithm}

Algorithm 2 has two loops. The first one (lines 2 to 11) is over literals.
The second one (lines 4 to 10) is over terms.

Clauses, literals, and terms are defined as follows:

\begin{enumerate}
\item Clause is an expression formed from a finite collection of literals (atoms or their negations).
\item Literal is an atomic formula (atom) or its negation.
\item A variable, a constant and an n-ary function symbol applied to n terms
\end{enumerate}

At line 6, OTTER computes paramodulants over current subterm lists.

\section{The hot list strategy}
The hot list strategy \cite{hot1} is one of the look-ahead strategies. Look-ahead strategies are designed to enable the program to evade many CPU hours to draw conclusions.
The conclusion to draw may require focusing on a retained clause.

\begin{definition}
\underline{Definition of the hot list strategy.}
The hot list strategy enables the program to specify the facts by revisiting the hot clause repeatedly in the context of completing the application of an inference rule.
\end{definition}

For implementing the hot list strategy, the main loop based on the given clause algorithm should be modified.
The main loop for inferring and processing clauses and searching for a refutation operates
mainly on the lists usable and SoS.

\begin{enumerate}
\item Choose appropriate $given\_clause$ in SoS;
\item Move $given\_clause$ from $list(SoS)$ to $list(usable)$
\item Infer and process new clauses using the inference rules set.
\item Newly generated clause must have the $given\_clause$.
\item Do the retention test on new clauses and append those to $list(SoS)$.
\end{enumerate}

Figure 2 shows the chart flow of modifying the main loop for the hot list strategy.
The hot list strategy is designed to make some set of clauses (hot lists) immediately considered with each newly
retained clause.
With the modification,
if the program passes the branch on the lower side of Figure 2, which is ``hotlist exists?'',
the paramodulation routine (para\_into) is immediately invoked in the post-process.

\begin{figure}
\centering
\includegraphics[width=1.05\columnwidth]{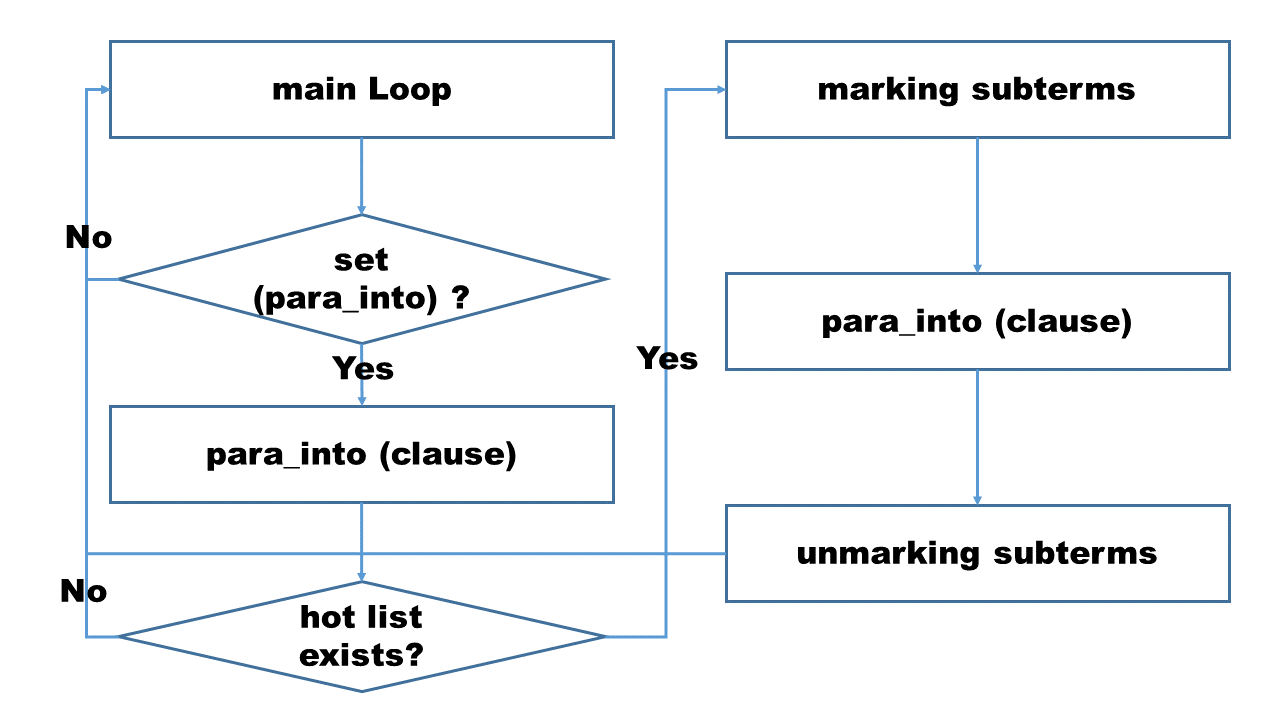}
\caption{Hot List Strategy by modifying main loop.}
\label{fig:image2}
\end{figure}

\section{Methodology}

\subsection{Setting Inferene Rule set}

As we discussed before, the basic inference mechanism of OTTER is based on
the given-clause algorithm.
Given-clause algorithm can be viewed as a simple implementation of the set of support strategy.
OTTER maintains four lists of clauses: usable, SoS, demodulator, and passive.
In our case, we cope with two kinds of clauses: usable and SoS.

Horizontal sliding from row[i] to row[i+1] is represented as follows.

\begin{verbatim}
list(usable).
EQUAL(l(hole,l(n(x),y)),l(n(x),l(hole,y))).
end_of_list.
\end{verbatim}

\begin{eqnarray}
\sigma = \left\{
\begin{array}{lllll}
1 & hole & 2 & 3 &\\
1 & 2 & hole & 3 &\\
\end{array}
\right\}
\end{eqnarray}

Vertical sliding from row[i] to row[i+4] is represented as follows.

\begin{verbatim}
list(usable).
EQUAL(l(hole,l(x,l(y,l(z,l(u,l(n(w),v)))))),
l(n(w),l(x,l(y,l(z,l(u,l(hole,v))))))).
end_of_list.
\end{verbatim}

\begin{eqnarray}
\sigma = \left\{
\begin{array}{lllllllll}
1 & hole & 2 & 3 & 4 & 5 & 6 & 7 \\
1 & 2 & 3 & 4 & 5 & hole & 6 & 7 \\
\end{array}
\right\}
\end{eqnarray}

Also, the initial state of the board is represented in the list(sos).

\begin{verbatim}
list(sos).
STATE(l(n(3),l(n(2),l(n(1),
l(end,l(n(8),l(n(4),l(n(7),
l(end,l(n(6),l(n(5),
l(hole,end)))))))))))).
end_of_list.
\end{verbatim}

\subsection{Generating puzzles (boards)}

In general, to check the solvability of N puzzles, the number of inversions of each number of N slots is calculated.

For example, if we have the board configuration board [2,3,6,1,7,8,5,4, hole]

(5,2,8,4,1,7, hole, 3,6), the number of inversions are as follows:

\begin{enumerate}
\item 2 precedes 1 - 1 inversions
\item 3 precedes 1 - 1 inversion
\item 6 precedes 1, 5, 4 - 3 inversions
\item 1 precedes none - 0 inversions
\item 7 precedes 5, 4 - 2 inversions
\item 8 precedes 5, 4 - 2 inversions
\item 5 precedes 4 - 1 inversions
\item 4 precedes none - 0 inversions
\end{enumerate}

Total inversions 1+1+3+0+2+2+1+0 = 10 (Even Number) So this puzzle configuration is solvable.
On the other hand, it is not possible to solve an instance of 8 puzzles if a number of inversions are odd in the input state.

\begin{algorithm}
\caption{Checking the solvability of N puzzles}
\label{alg1}
\begin{algorithmic}[1]
\renewcommand{\algorithmicrequire}{\textbf{Input:}}
\renewcommand{\algorithmicensure}{\textbf{Output:}}
\REQUIRE Board[$ x_1, x_2, ..., x_n, hole $]
\ENSURE SOLVABLE or UNSOLVABLE
\STATE $ Board[X | XS] = Board[x_1, x_2, ..., x_n, hole] $
\WHILE{ $XS$ in $Board[X l XS] $ is empty}
\FOR {$i$ in $XS$}
\STATE statements..
\IF {($ X \ne XS[i] $)}
\STATE $ counter[i]++$
\ENDIF
\ENDFOR

\ENDWHILE

\STATE $ line = check(Board[...] \subseteq hole) $

\STATE $ sum = 0 $
\FOR {$i$ to $n$}
\STATE $ sum += counter[i]$
\ENDFOR

\IF {($ line + sum \% 2 == 0 $)}
\STATE $ flag =$ SOLVABLE
\ELSE
\STATE $ flag =$ UNSOLVABLE
\ENDIF

\end{algorithmic}
\end{algorithm}

Algorithm 3 shows the procedure for checking the solvability of N puzzles.
At lines 2 to 9, the number of inversions of each slot is counted.
These figures are counted up at lines 11 to 14.
Finally, the sum is checked if it is an even or odd number at lines 15 to 19.

\section{Experimental results}

\subsection{Generating puzzles (boards)}

In the experiment, we have generated 500 sliding puzzles with size 8 * 8.
All generated configurations of 8 puzzles are solvable.
For each puzzle, we have measured the number of generated clauses with the procedures shown in Algorithm 2.
For simplicity, we have generated the configuration of the first 8 slots with random integers ranging from 1 to 8 and fixed 9th slot to hole.

\subsection{Counting clauses}

Algorithm 4 shows the brief description of the modified given clause algorithm for counting the generated clauses.

\begin{algorithm}
\caption{Incrementing the number of generated clauses}
\label{alg1}
\begin{algorithmic}[1]

\WHILE{given clause is NOT NULL}
\STATE $ index\_lits\_clash(giv\_cl); $
\STATE $ append\_cl(Usable, giv\_cl); $

\IF{$ splitting() $ }
\STATE $ possible\_given\_split(giv\_cl); $
\ENDIF
\STATE{infer\_and\_process(giv\_cl);}
\STATE{giv\_cl = extract\_given\_clause();}
\STATE{track(the\_number\_of\_generated\_clauses);}

\ENDWHILE
\end{algorithmic}
\end{algorithm}

At line 9, the number of generated clauses is incremented.
After line 8 of picking up the clause from a set of support, we can record the current size of the set of support.
By doing this, we can obtain the plot with \# puzzles and the number of generated clauses of the Y-axis, as shown in the next section.

Table I shows the numerical results of solving 500 puzzles randomly generated.
The number of generated clauses with paramodulation ranges from 510 (1,3,5,4,6,8,7,2,hole - easiest) to 188,610 (6,2,7,3,4,5,8,1,hole - the most difficult).
In the view of complexity of reasoning process, the configuration [ (\#295 1,3,5,4,6,8,7,2,hole)] is 369.82 times harder to solve than
the configuration [\#124 (6,2,7,3,4,5,8,1,hole)].

\begin{table}[htbp]
\begin{center}
\caption{Initial board states and the complexities of paramodulation}
\begin{tabular}{{|l|l|l|}} \hline
board No & Initial state & clauses generated \\ \hline \hline
\#295 & 1,3,5,4,6,8,7,2,hole & 510 (easiest) \\ \hline
\#340 & 8,1,3,4,2,5,7,6,hole & 918 \\ \hline
\#294 & 2,3,5,1,6,8,7,4,hole & 1,362 \\ \hline
\#86 & 8,7,6,4,5,2,3,1,hole & 188,475 \\ \hline
\#124 & 6,2,7,3,4,5,8,1,hole & 188,610 (the most difficult)\\ \hline
\end{tabular}
\end{center}
\end{table}

\subsection{Heat flow}

In nature, sliding puzzles are two-dimensional, even if the sliding is facilitated by encaged marbles or three-dimensional tokens.
We define the heat flow in paramodulation as follows.

\begin{definition}
\underline{Definition of heat flow.}
Heat flow makes the reasoning program consider the hot list immediately with vertical and horizontal permutation.
\end{definition}

Horizontal heat flow is set by the hot clause as follows:

\begin{verbatim}
list(hot).
EQUAL(l(hole,l(n(x),y)),l(n(x),l(hole,y))).
end_of_list.
\end{verbatim}

Also, vertical heat flow is set by the hot clause as follows:

\begin{verbatim}
list(hot).
EQUAL(l(hole,l(x,l(y,l(z,l(u,l(n(w),v)))))),
l(n(w),l(x,l(y,l(z,l(u,l(hole,v))))))).
end_of_list.
\end{verbatim}

\begin{table}[htbp]
\begin{center}
\caption{The number of clauses generated by vertical/horizontal heat flow}
\begin{tabular}{{|l|l|l|l|}} \hline
Board No & Initial state & vertical & horizontal \\ \hline \hline
\#295 & 1,3,5,4,6,8,7,2,hole & 254 & 388 \\ \hline
\#340 & 8,1,3,4,2,5,7,6,hole & 502 & 725 \\ \hline
\#294 & 2,3,5,1,6,8,7,4,hole & 1,290 & 1,964 \\ \hline
\#86 & 8,7,6,4,5,2,3,1,hole & 110,423 & 100,803 \\ \hline
\#124 & 6,2,7,3,4,5,8,1,hole & 96,809 & 100,824 \\ \hline
\end{tabular}
\end{center}
\end{table}

\begin{figure}
\centering
\includegraphics[width=0.8\columnwidth]{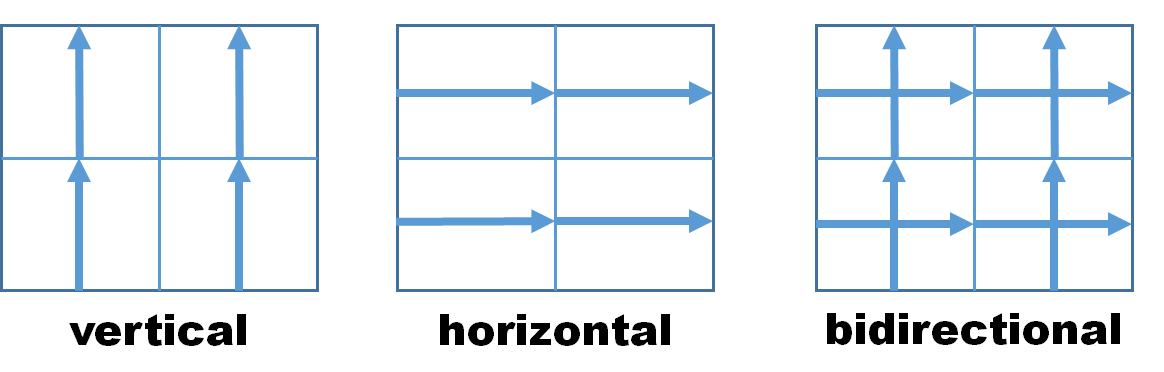}
\caption{Heat flow in paramodulation.}
\label{fig:image2}
\end{figure}

\begin{figure}
\centering
\includegraphics[width=1.0\columnwidth]{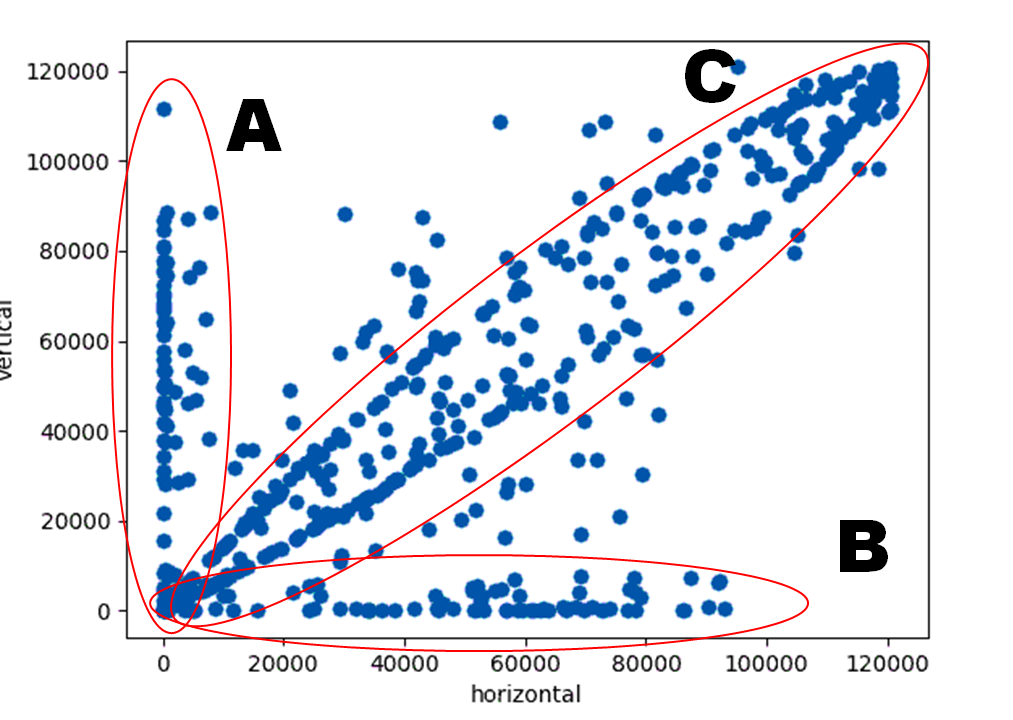}
\caption{2D scatter diagram of the number of clauses generated in solving 500 puzzles. }
\label{fig:image2}
\end{figure}

Figure 4 shows the number of clauses generated in solving 500 puzzles.
Figure 4 has 500 points of the initial state of the board.
The X-axis is the number of clauses generated with vertical heat flow.
Y-axis is the number of clauses generated with horizontal heat flow.
That is, there are 500 points of boards with point (x,y) where x is the number of clauses
generated with horizontal heat flow and y is the number of clauses in vertical heat flow.
For example, the points \#295 have the values (388,254) as shown in Table II.

In Figure 4, we distiguish three areas among 500 points.

\begin{enumerate}
\item Area A: The boards are affected by vertical heat flow.
\item Area B: Horizontal heat flow are effective on the boards.
\item Area C: Both vertical and horizontal heat flow have effects on the boards.
\end{enumerate}

Figure 5 is a 3D scatter diagram of 500 boards.
Points in Area C in Figure 4 are also plotted in Figure 5.
We see the bulk of points (Area D) in the lower side of the figure (seemingly with little effect of bidirectional heat flow).

\begin{figure}
\centering
\includegraphics[width=1.0\columnwidth]{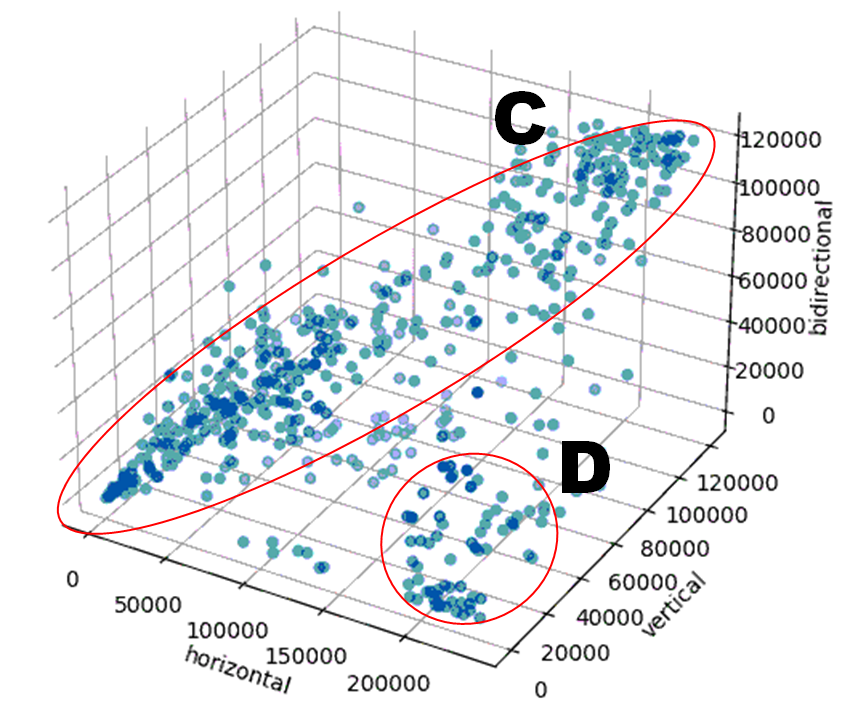}
\caption{3D scatter diagram of the number of clauses generated in solving 500 puzzles. }
\label{fig:image2}
\end{figure}


\section{Related work}
Historically, Noyes Chapman invented the oldest type of sliding puzzle, which is the fifteen puzzle in 1880.
Folklore tells us that in 1886, puzzle master Sam Loyd offered a one-thousand dollar prize if anyone could swap tile 14 and 15 and return the other tiles to their original slots.
Archer \cite{Archer} firstly discusses an algorithmic analysis of 15 puzzles.
In \cite{Archer}, a summary of all possible permutations of slots attained by moving the black block from cell i to cell j affecting the permutation of $ \sigma_i,j $.
Howe \cite{Howe} proposes two approaches in the two kinds of viewpoints: the properties of permutations and graph theory.
Calabro \cite{Calabro} proposes $ O(n^2) $ time algorithm for deciding the time when the initial configuration of the n * n puzzle game is solvable.


Paramodulation originated as development of resolution \cite{robinson}, one of the main computational methods in first-order logic, see \cite{Bachmair}.
For improving resolution-based methods, the study of the equality predicate has been particularly important since reasoning with equality
is well-known to be of the great importance of mathematics, logic, and computer science.
Ando et al. \cite{ando} propose a measurement of the complexity of sliding block puzzles using paramodulation.

\section{Conclusion}

In this paper, we have presented the new novel experiments of the complexity of sliding block puzzles based on the concept of heat flow in paramodulation. 
Heat flow is set by the hot list with vertical and horizontal permutation.
In the experiment, we have generated 500 * 8 puzzles to calculate the number of clauses generated by vertical and  horizontal heat flow in the board.
We have obtained some curious results. To name a few, board \#295 (1,3,5,4,6,8,7,2,hole) turned out to be easiest with the 2D coordinate (388, 254).
Board \#124 (6,2,7,3,4,5,8,1,hole) is the most difficult with the 2D coordinate (96,809 100,824).
Also, we have distinguished three areas in 500 points.
For one possible further work, we are aiming to leverage this research for the hybrid of algorithmic modules providing formal reasoning, search, and abstraction capabilities with geometric modules.

\end{document}